\documentclass[aps,prx,twocolumn,showpacs,superscriptaddress,groupedaddress]{revtex4-1}  
\pdfoutput=1
\usepackage{graphicx}
\usepackage{amsmath}

\def\a{\alpha}
\def\b{\beta}

\def\r{\mathbf{r}}
\def\bnabla{\boldsymbol{\nabla}}

\global\long\def\n{\mathbf{\hat{n}}}
\global\long\def\np{\mathbf{\hat{n}}_{\perp}}
\def\j{\mathrm{J}}

\begin{document}
\title{Geometric frustration and compatibility conditions for two dimensional director fields}
\date{\today}

\author{Idan Niv}
\affiliation{Department of Physics of Complex Systems, Weizmann Institute of Science, Rehovot 76100, Isreal}
\author{Efi Efrati}
\email{efi.efrati@weizmann.ac.il}
\affiliation{Department of Physics of Complex Systems, Weizmann Institute of Science, Rehovot 76100, Isreal}


\begin{abstract}
The uniform director field obtained for the nematic ground state of the hard-rod model of liquid crystals in two dimensions reflects the high symmetry of the constituents of the liquid \cite{Ons49}; It is a manifestation of the constituents' local tendency to avoid splaying and bending with respect to one another. 
In contrast, bent-core (or banana shaped) liquid-crystal-forming-molecules locally favor a state of zero splay and constant bend. However, such a structure cannot be realized in the plane \cite{Mey76} and the resulting liquid-crystalline phase is frustrated and must exhibit some compromise of these two mutually contradicting local intrinsic tendencies. The generation of geometric frustration from the intrinsic geometry of the constituents of a material is not only natural and ubiquitous but also leads to a striking variety of morphologies of ground states \cite{Gra16,KS01} and exotic response properties \cite{LS16}.

	In this work we establish the necessary and sufficient conditions for two scalar functions, $s$ and $b$ to describe the splay and bend of a director field in the plane. We generalize these compatibility conditions for geometries with non-vanishing constant Gaussian curvature, and provide a reconstruction formula for the director field depending only on the splay and bend fields and their derivatives. Last, we discuss optimal compromises for simple incompatible cases where the locally preferred values of the splay and bend cannot be globally achieved.
\end{abstract}
\maketitle


The curved geometry of bent-core liquid crystal forming molecules is naturally associated curving of the director field along the long axis of the molecule, while minimizing the volume per particle locally favors packing the molecule parallel to one another across the long direction rendering them uniformly spaced. Mathematically these tendencies are associated with a preferred state of a constant non-zero bend and vanishing splay, respectively \footnote{see equation \eqref{eq:FrankConstBend} and the 2D polar limit of \cite{PSB+16}.}.
However, as is well known \cite{Mey76} (and derived below), there exists no director field in the plane that displays a vanishing splay and constant non-zero bend. The inaccessibility of this uniformly bent ground state is, at least partially, responsible for the plethora of locally stable morphologies observed for bent core liquid crystals including columnar, smectic and polar chiral phases \cite{LR02,TT06,SDS13}. 
To adequately describe the order parameter of such liquid crystals one has to employ a third rank tensor  in addition to the vector and the second-rank tensor order parameters \cite{LR02}. However, much of the rich structure of these phases, and in particular the notion of geometric frustration already arises when considering the director field alone.

The case of bent core liquid crystals constitutes a particular example of what has been recently termed frustrated assemblies \cite{Gra16} and has seen numerous experimental realizations using colloidal particles \cite{GBZ+12,MPNM14}. 
While one could pursue the individual equilibria shapes through direct molecular simulation, unravelling the general principles governing the assembly of such frustrated structures requires a coarse grained continuous geometric description. Such a geometry based description could identify universal modes of frustration associating seemingly disparate systems with the same type of geometric frustration, and thus with similar morphologies and solutions. In general, in such a coarse grained description the geometric frustration is associated with a mismatched geometric charge \cite{MPNM14,SG05}, and one of the central challenges in formulating such a description is to properly identify the geometric charge associated with the mutually incompatible local preferences of the building blocks of the assembly.
\begin{figure}[h]
\centerline{\includegraphics[width=.45\textwidth]{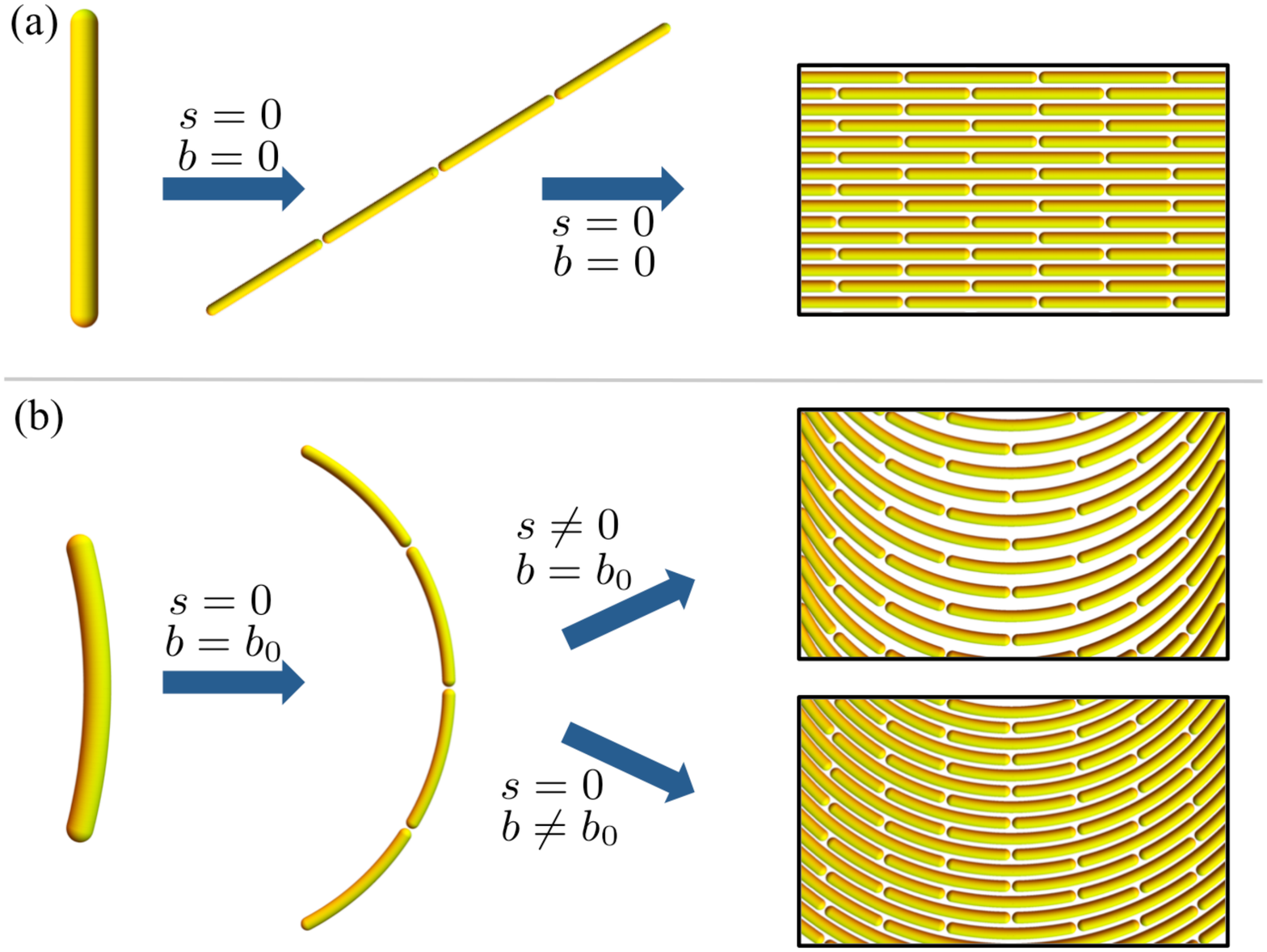}}
\caption{
(a) The intrinsic geometry of a straight rod favors a state with vanishing splay and bend. This can be realized by a uniform director field.
(b) The intrinsic geometry of a banana shaped rod favors a state of vanishing splay and constant bend. This state, however, cannot be achieved. Two possible compromises are presented; a state with vanishing splay (equidistant layers) necessitates bending variations as observed in the concentric circular director of the bottom panel. A constant bend state results inevitably in non-vanishing splay as seen in the non-uniform spacing in the top panel.}\label{fig:schematic}
\end{figure}

In two dimensional Riemannian geometry only one such geometric charge exists and corresponds to the integrated Gaussian curvature. Frustration in this particular case occurs when the assembling constituents are attributed one value of Gaussian curvature, while the two dimensional space in which they assemble is associated with a different valued Gaussian curvature.  For a constant curvature difference, as the assembly grows it accumulates the geometric mismatching charge proportionally to its area. However, the energy associated with this increasing charge as well as the deformations it gives rise to typically grow much faster than the area resulting in size limitation of the assembled structure favoring narrow assemblies of large aspect ratio \cite{ESK13,SG05,MPNM14}. 

Focusing on director fields in two dimensions we aim to formalize the notion of frustration for such systems and  address the possible ways to resolve this frustration. In this work we provide an explicit formula for the Gaussian curvature associated with given splay and bend fields, thus mapping the geometrically frustrated problem of bent-core liquid crystals (or any other local splay/bend tendency) in two dimensions to the much studied realm of optimal embedding of manifolds of mismatched Gaussian curvatures.  A problem encountered in elasticity \cite{ESK09,AEKS11,ESK13}, crystal growth of curved surfaces \cite{SG05,MPNM14} and the bundling of twisted filaments \cite{HBBG16,BG12}.

\section{Pre-Compatibility conditions in the plane}
%
We start by considering a director field $\n$ in the plane, setting $\n=(\cos(\theta),\sin(\theta))$ and $\np=(-\sin(\theta),\cos(\theta))$.
In this case the splay, $s$, and bend, $b$, of the director field may be related to the directional derivatives of the angle $\theta$ that the director forms with the $x$-axis. 
\begin{equation}
\begin{aligned}
s=&\bnabla\cdot\n=\theta_{y}\cos\theta-\theta_{x}\sin\theta=\np\cdot\bnabla\theta,\\
b=&\left|\n\times\bnabla\times\n\right|=\left|\n\cdot\bnabla\n\right|=\left|\theta_{x}\cos\theta+\theta_{y}\sin\theta\right|\\
=&\left|\n\cdot\bnabla\theta\right|.
\end{aligned}
\label{eq:s&b}
\end{equation}
When the bend is bound away from zero we may replace the non-negative definition of the bend used above with the signed bend
$b=\n\cdot\bnabla\theta$. For simplicity we assume throughout what follows that the bend is non-negative.

The above relations may be inverted to give the derivatives of $\theta$ as a function of the bend and splay. 
\begin{equation}
\bnabla\theta=b\n+s\np=\begin{pmatrix}\cos\theta & -\sin\theta\\
\sin\theta & \cos\theta
\end{pmatrix}\begin{pmatrix}b\\
s
\end{pmatrix}.
\label{eq:gradtheta}
\end{equation}
Equations \eqref{eq:gradtheta} may be considered as two first order non-linear partial differential equations for $\theta$ given the splay and bend fields. These equations may be integrated to give the director only if the mixed second derivatives
of $\theta$ commute, i.e. $\partial_{x}\partial_{y}\theta=\partial_{y}\partial_{x}\theta$. In other words we verify that the field $\bnabla\theta$ is indeed conservative. This condition may be compactly written as 
\[
\bnabla\times (b\n+s\np)= \bnabla\cdot\left(s\n-b\np\right)=0.
\]
When the definition of the bend and splay as expressed in \eqref{eq:gradtheta} are substituted into this equation we obtain the pre-compatibility equation:
\begin{equation}
b^{2}+s^{2}+\n\cdot\bnabla s-\np\cdot\bnabla b=0.
\label{eq:flatprecompatibility}
\end{equation}
This equation provides a necessary condition for the existence of a director field with bend $b$ and splay $s$.
It is called pre-compatibility as it contains explicitly (through $\n$ and $\np$) the function $\theta$ whose existence is sought. Nonetheless it already captures the textbook version of incompatible fields \cite{Mey76}. Considering only uniform fields, then all gradients in equation \eqref{eq:flatprecompatibility} vanish and we obtain $b^{2}+s^{2}=0$, implying that in this case only the trivial solution $s=b=0$ is admissible. 

\section{Pre-Compatibility in non-Euclidean geometries}
\label{sec:Pre-comp-NE}
\begin{figure}[t]
\centering
\includegraphics[width=.8\linewidth]{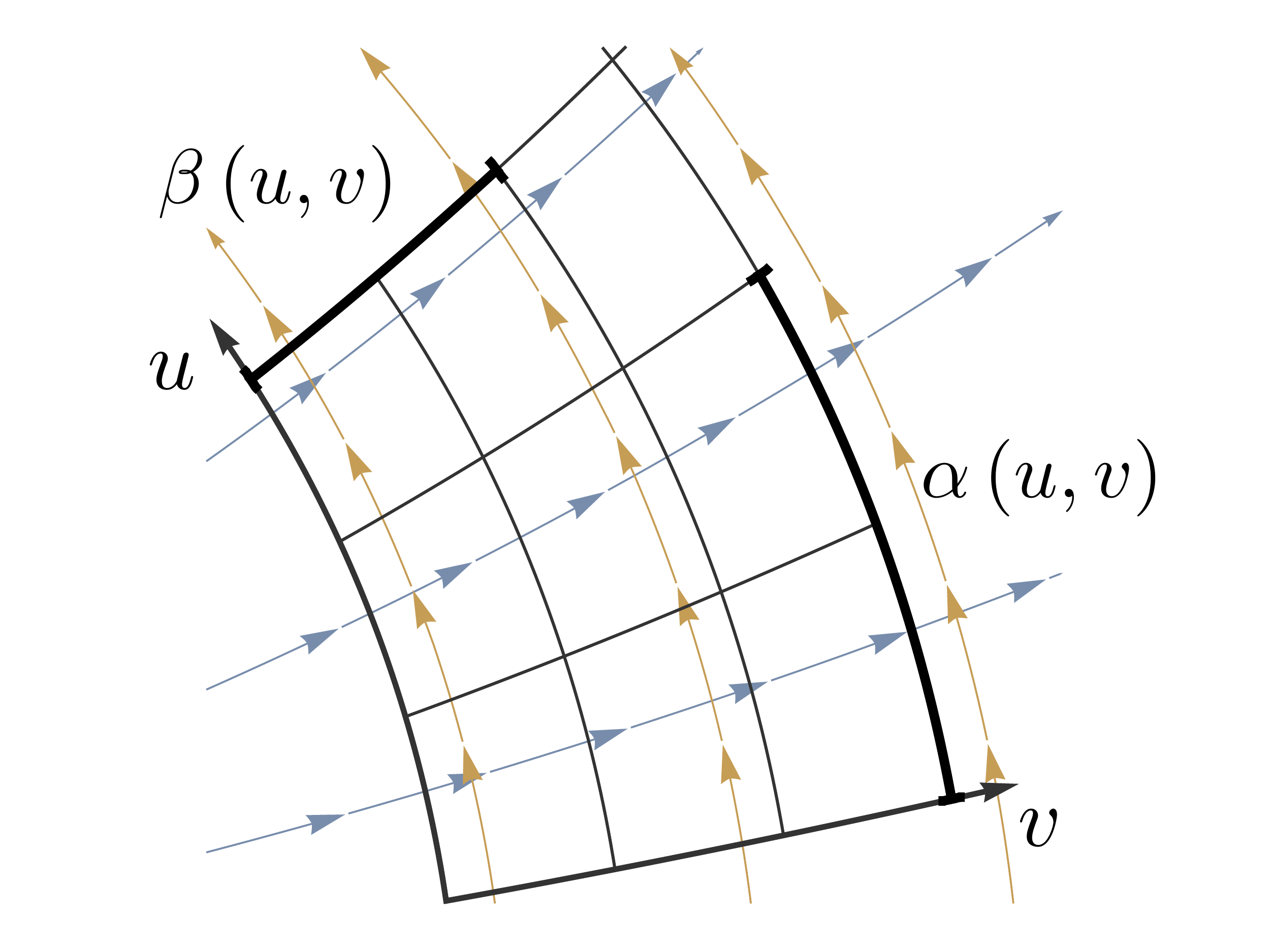}
\caption{Director coordinate system. We construct the coordinate system by following the integral line of the director field this is the $u$ axis. As we go on $u$ we can draw lines going on the normal direction which point in the $v$ direction. Moving a $v$ line along the $u$ direction we see that its 2 sides could move at different paces. The velocity of the point $(u,v)$  in this motion is $\alpha(u,v)$. We rescale the $v$ coordinate such that $\alpha(u=0,v)=1$. We repeat the process for the other direction with $\beta$.
}\label{fig:grid}
\end{figure}

We now come to consider the more general setting in which the director field is defined on a surface $S$ (not necessarily Euclidean). There is a natural local orthogonal coordinate system, $(u,v)$, induced by the director field in which the $u$-parametric-curves (along which $v=const$) are everywhere tangent to $\n$, and the $v$-parametric-curves (along which $u=const$) are everywhere tangent to $\np$. See figure \ref{fig:grid} and appendix \ref{app:coordinates} for more details on the explicit construction.

With respect to the parametrization of $S$ given by $\r(u,v)$, we have
\begin{equation}
\partial\r/\partial u\equiv\r_{u}=\alpha\n,\quad\text{and}\quad\partial\r/\partial v\equiv\r_{v}=\beta\np.
\label{eq:parametrization}
\end{equation}
There is a gauge freedom for each of the arc length functions which we eliminate by setting $\alpha(u,v=0)=1$ and $\beta(u=0,v)=1$. Up to the choice of the location of the origin, this results in a unique Riemannian metric reading 
\begin{equation}
d\sigma^{2}=\alpha^{2}du^{2}+\beta^{2}dv^{2}.
\label{eq:metric}
\end{equation}
With respect to this parametrization the directional derivatives along the director and its normal are associated with the arc-length differentiation along the coordinates $u$ and $v$ respectively:
\begin{equation}
\label{eq:derivatives}
(\n\cdot\bnabla) f=\left.\frac{1}{\alpha}\frac{\partial f}{\partial u}\right|_{v}\,, \quad
\text{and}\quad
(\np\cdot\bnabla) f=\left.\frac{1}{\beta}\frac{\partial f}{\partial v} \right|_{u}\,.
\end{equation}

We note that for a general director field on a curved surface one could define two distinct bend measures. The first, purely intrinsic and measures the geodesic curvature of the director's integral curve on the surface, $\kappa_{g}$. The second measures the curvature of the director field integral curve in three dimensions, i.e. $|\partial_{\sigma}\partial_{\sigma}\boldsymbol{\gamma}|=\kappa $, where $\sigma$ denotes the arc-length parameter along the integral curve $\boldsymbol{\gamma}$. This measure depends on the specific embedding of the surface and relates to the geodesic curvature through $\kappa^{2}=\kappa_{g}^{2}+\kappa_{n}^{2}$, where $\kappa_{n}$ is the normal curvature of the surface along the direction of $\n$. For the case of a director field in the plane, these two measures coincide. Seeking to obtain a strictly intrinsic relation between the splay and bend, in what follows we identify the bend with the first, purely intrinsic definition of the bend. In agreement with the flat case the splay is defined through $s=\bnabla\cdot\n$. Applied to the metric \eqref{eq:metric} these definitions yield
 \begin{equation}
s= \frac{\beta_{u}}{\alpha\beta},\quad \text{and}\quad
b= - \frac{\alpha_{v}}{\alpha\beta}.
\label{eq:s&b2}
\end{equation}
One can calculate the Gaussian curvature, $K$, of the surface $\r$ directly from its metric \cite{Str88} through
\begin{equation}
\begin{aligned}K & =-\left(-\frac{\alpha_{v}}{\alpha\beta}\right)^{2}-\left(\frac{\beta_{u}}{\alpha\beta}\right)^{2}-\frac{1}{\alpha}\partial_{u} \frac{\beta_{u}}{\alpha\beta}+\frac{1}{\beta}\partial_v \frac{\alpha_{v}}{\alpha\beta}=\\
 & =-b^{2}-s^{2}-\frac{1}{\alpha} s_u+\frac{1}{\beta} b_v=\\&=-b^{2}-s^{2}-\n\cdot\bnabla s+\np\cdot\bnabla b.
\end{aligned}
\label{eq:precompatibility}
\end{equation}
Similarly to equation \eqref{eq:flatprecompatibility} this equation explicitly contains the director $\n$ whose existence is sought and thus is similarly termed the pre-compatibility condition. It constitutes a necessary condition for the existence of an embedding of director field $\n$ with splay $s$ and intrinsic bend $b$ on a surface of Gaussian curvature $K$. We can now easily identify \eqref{eq:flatprecompatibility} with the particular case where $K=0$.
 
Considering again the special case of uniform splay and bend fields, where all gradient vanish, the explicit dependence of the precompatibility condition on $\n$ is eliminated and we obtain the compact form
\begin{equation}
b^{2}+s^{2}=-K.
\label{eq:comp_unifrom}
\end{equation}
The case for $K=0$, solved in the previous section addmitted only the trivial solutions $s=b=0$. For dome-like surfaces, where $0<K$, even the trivial solution is not admissible and we obtain that this geometry cannot support any director field with constant bend and splay. Last, for saddle-like surfaces, where $K<0$, non-trivial solutions with constant bend and splay exist, provided $|K|$ is large enough such that the splay obtained through $s=\sqrt{|K|-b^{2}}$ is real. 

\section{The Reconstruction formula and Eulerian compatibility conditions}
The  precompatibility condition \eqref{eq:precompatibility} constitutes a necessary condition for the existence of a director field with bend $b$ and splay $s$. Its satisfaction, however, does not necessarily imply the existence of such a director, as in obtaining this form we have made explicit use of the definition of splay and bend fields:
\begin{equation}
b=|\n\cdot\bnabla\n|,\qquad s= \bnabla\cdot\n.
\label{eq:defs&b}
\end{equation}
We next rewrite the precompatibility condition as an explicit reconstruction formula for $\n$ as a function of the splay and bend fields and their gradients. However, this reconstruction will be meaningful only if it trivially satisfies the above definition for the bend and splay. 

We first seek to eliminate the unit vector $\np$ from \eqref{eq:precompatibility}. To do so we define the operator $\j$ to be a $\pi/2$ rotation such that $\j\n=\np$, and identify that $\np\cdot\bnabla b=\j\n\cdot\bnabla b=-\n\cdot \j\bnabla b$. We thus may reinterpret \eqref{eq:precompatibility} as the projection of the director on the vector $\mathbf{V}=\bnabla s+\j\bnabla b$, .i.e. $\n\cdot\mathbf{V}=\n\cdot\left(\bnabla s+\j\bnabla b\right)=-b^{2}-s^{2}-K$. 
We now define the two unit vectors
\begin{equation}
\hat{\mathbf{V}}=\frac{\bnabla s+\j\bnabla b}{\left|\bnabla s+\j\bnabla b\right|}
\, , \quad 
\hat{\mathbf{V}}_\perp=\j\hat{\mathbf{V}}=\frac{-\bnabla b+\j\bnabla s}{\left|\bnabla s+\j\bnabla b\right|}.
\label{eq:Vperp&V}
\end{equation}
Except for the case where $\mathbf{V}=\mathbf{0}$, these two unit vectors span the local tangent space and thus may be used to express $\n$ \footnote{If the splay and bend fields satisfy $\bnabla s+\j\bnabla b=0$, the vector $\mathbf{V}$ cannot be defined. However, for this case we obtain that $K=-s^2-b^2$ identically, which for constant Gaussian curvature can be easily shown to lead only to the uniform solution case. See appendix \ref{Appendix:Vis0}.}. Equation \eqref{eq:precompatibility} provides the coefficients explicitly:
\begin{equation}
\n=-\frac{b^{2}+s^{2}+K}{\left|\bnabla s+\j\bnabla b\right|}\hat{\mathbf{V}} \pm \sqrt{1-\left(\frac{b^{2}+s^{2}+K}{\left|\bnabla s+\j\bnabla b\right|}\right)^{2}}\hat{\mathbf{V}}_{\perp}.
\label{eq:directorReconstruct}
\end{equation}
Note that the sign ambiguity above is artificial as only one of the two branches yields the correct reconstructed values for $b$ and $s$. However, in different cases different values of the sign need to be assigned, and the sign is generally chosen to assure continuity for the calculated fields. 

The compatibility conditions are now obtained by explicitly substituting \eqref{eq:directorReconstruct}  into the definition of the splay and bend \eqref{eq:defs&b}. This results in two non-linear, second order partial differential equations for $s$ and $b$. The satisfaction of these equations assures the existence of a director field $\n$ for which the splay and bend are given respectively by $s$ and $b$. In this case the director is given explicitly by the reconstruction formula 
\eqref{eq:directorReconstruct} which depends only on the splay and bend fields and their gradients. This is in contrast with the integral formulae for reconstructing a vector field from knowledge of it curl and divergence fields as given by the 
Helmholtz theorem (see \cite{Bha13}). While the constraint for the unit length of the director field, $|\n|=1$ resulted in the non-trivial compatibility conditions, it also allowed the reconstruction of the director without resorting to the use of integral formulas. Instead we can calculate the director from knowledge of the curl and divergence fields and their gradients alone. 

The actual form of the compatibility conditions following this Eulerian formulation is very cumbersome and opaque, and only rarely admits an explicit solution. It may be useful for the specific cases where the bend and splay fields are given explicitly and one only seeks to find if they indeed describe a valid director field. In order to use the compatibility conditions to explicitly obtain a solution we formulate them using the director's natural coordinate system.


\section{Lagrangean formulation of compatibility conditions}
The vectorial formulation of the compatibility conditions appearing in the previous section can be applied to any set of curvilinear coordinates parameterizing the surface of Gaussian curvature $K$. However, if the coordinates are to follow the directions of the director field, as in figure \ref{fig:grid}, then knowing the way these coordinate behave in space is equivalent to solving for the director. Such a Lagrangean description of the splay and bend fields is of great interest as it provides a natural parameterization for the director and may better capture the intrinsic nature of the attempted splay and bend fields. In this parametrization, however, the compatibility conditions assume a slightly different form.

If a metric $d\sigma^{2}=\a^{2}du^{2}+\b^{2}dv^{2}$, satisfies \eqref{eq:precompatibility}, meaning its Gauss curvature as calculated from the metric matches the curvature of its embedding space, $K$, then one can find an isometric embedding of this metric that is unique up to rigid motions. The unknowns in the metric are related to the splay and bend through equation \eqref{eq:s&b2}. We use the compatibility condition to relate $\a$ and $\b$ and then eliminate one of them in \eqref{eq:s&b2}. For example we may use 
\begin{equation}
\begin{aligned}\b=\frac{ b_v}{K+b^{2}+s^{2}+\frac{1}{\alpha} s_u}
\end{aligned}
\label{eq:betaSubstitution}
\end{equation}
to express $\a_{u}$ and $\a_{v}$ as a function of $\a,\,b$ and $s$ \footnote{The above elimination assumes that either $s_{u}\ne 0$ or $b_{v}\ne 0$. The where case both vanish  corresponds to the $\mathbf{V}=0$ discussed earlier and in appendix \ref{Apendix:Vis0}}. Here too the obtained equations may be interpreted as first order PDE defining $\a$ that permit a solution only if $\partial_{v}\a_{u}=\partial_{u}\a_{v}$. Whenever this solvability condition is not trivially satisfied we can obtain from it a cubic polynomial equation for $\a$. The solution to this polynomial equation,
together with $\b$ defined through \eqref{eq:betaSubstitution} are substituted to equation \eqref{eq:s&b2} providing the compatibility conditions for the $s$ and $b$ fields in the natural $u$ and $v$ coordinate system.

\section{Using the compatibility conditions to obtain explicit solutions}
We now come to exploit the compatibility conditions in order to obtain the equilibrium solution for the incompatible case of bent core liquid crystals. For such systems, the nematogen favors a state of vanishing splay and constant bend. As was previously shown, for example in equation \eqref{eq:comp_unifrom}, such a state cannot exist in the plane. Thus every realized configuration will inevitably contain some compromise between the mutually contradicting bend and splay tendencies. The specific equilibrium will naturally depend on both the dimensions of the region under consideration and on the relative magnitude of the Frank energy constants. For a positively defined bend, $b$, the Frank free energy for a 2D bent-core liquid crystal is given by \footnote{This expression for the energy can be deduced by taking the two dimensional and polarized limit of the polar nematic theory of bent core liquid crystals, as for example appears in \cite{PSB+16}, and identifying the polar vector with the perpendicular to the director, $\mathbf{p}\parallel \np$. For non-positively defined $b$, it is natural (albeit less convenient) to formulate the bend term using squares: $(b^{2}-\bar{b}^{2})^{2}$.\cite{leo}}

\begin{equation}
	E=\int\left[K_1 s^2+K_3 \left(b-\bar{b}\right)^2\right]dA.
	\label{eq:FrankConstBend}
\end{equation}
\subsection{Splay free director field} 
We begin by examining the case where $K_{3}\ll K_{1}$. In this limiting case one expects that whenever possible the solution will display vanishing splay, and minimize the remaining energy with respect to the bend among all such splay free solutions. Setting $s=0$, equation \eqref{eq:s&b2} reads $\partial_u \beta=0$. We are allowed to set the initial value of $\beta$ along the $u=0$ curve, as explained in Figure \ref{fig:grid}, $\beta(u=0,v)=1$, which leads after integration to $\beta(u,v)=1$. Thus, the natural coordinate system forms a semi-geodesic parametrization for the plane \cite{Str12}.
When substituted into \eqref{eq:precompatibility} we obtain for the bend the equation $b_v=b^2+K$,
which for the case $K=0$ yields 
\begin{equation}
b=\frac{b_{0}}{1-b_{0} v}.
\end{equation}
The metric coefficient is given by
$\a=b_{0}/{b}=1-b_{0} v$, corresponding to a director oriented along concentric circles as appearing in Figure \ref{fig:schematic}(ii).
The function $b_{0}$ is in turn chosen such as to minimize the energy \eqref{eq:FrankConstBend}. 
Considering a finite narrow region in space, $0\le u \le L$ and $-w/2 \le v \le w/2$, where $w/L\ll 1$ and $w \bar{b}\ll1$ this minimization yields 
$ b_{0}(u)\approx\bar{b}\left(1-w^{2}\bar{b}^{2}/6\right)$.
The energy associated with this state of non uniform bend to leading order in the width reads
\[
E\approx K_{3}\frac{\bar{b}^{4}w^{3}L}{12}\left(1- \frac{\bar{b}^{2}w^{2}}{3} \right).
\]
Note that this energy grows super-extensively for domains of constant aspect ratio; if $w L =A$ and $w/L$ is held constant then to leading order the energy per unit area grows according to $E/A\propto A$. However, if the width is held constant and only the length is changed then $E/A$ is also constant. Such behavior is very typical of geometrically frustrated assemblies and often leads to filamentation \cite{HBBG16,ESK13,MPNM14,Gra16,SG05} .

\subsection{Constant bend director field} 
We now consider the opposite limit in which $K_{1}\ll K_{3}$. Similarly we now expect that, whenever possible, the system will assume a state where $b=\bar{b}$ and minimize the remaining energy with respect to the splay among all such constant bend solutions. Setting $\bar{b}=b=-\a_{v}/\a\b$ we obtain from equations \ref{eq:s&b2} and \ref{eq:precompatibility}
\[
\partial_{u}\bigl(\b^{2} (s^{2}+\bar{b}^{2}) \bigr)=0,\qquad \bar{b}\a =-\partial_{u}\arctan(s/\bar{b}).
\]
Upon substiting $s=\bar{b}\tan(\Theta)$ we obtain $
\b= c(v) \cos(\Theta)$, for some arbitrary function $c(v)$, and
$\bar{b}=-\frac{1}{\a}\partial_{u}\Theta$. The compatibility conditions reduce to 
\[
\partial_{u}\bigl(\Theta_{v}+\bar{b} c(v)\sin(\Theta) \bigr)=0.
\]
Integrating the equation above produces an arbitrary function of $u$.
Considering a narrow ribbon $-w/2\le u \le w/2$ and $-L/2 \le v \le L/2$ where again $w/L\ll 1$ and $w \bar{b}\ll1$, we may set the arbitrary function to zero and also assume that $s(u=0,v)=0$ \cite{bote2}. 
We are thus led to
\[
 \tan(\Theta/2)=A(u)B(v),
\]
where setting the initial values of the metric coefficients $\a(u,v=0)=1$ and $\b(u=0,v)=1$ yields $A(u)=-\tan(\bar{b} u/2)$ and $B(v)=\exp(-\bar{b} v)$. 
Similarly to the splay free case, to leading order the Franck energy scales as
 $E\approx K_{1}\frac{\bar{b}^{4}w^{3}L}{12}$.

\subsection{Equal Frank coupling constants}
\begin{figure}[h]
\centerline{\includegraphics[width=.35\textwidth]{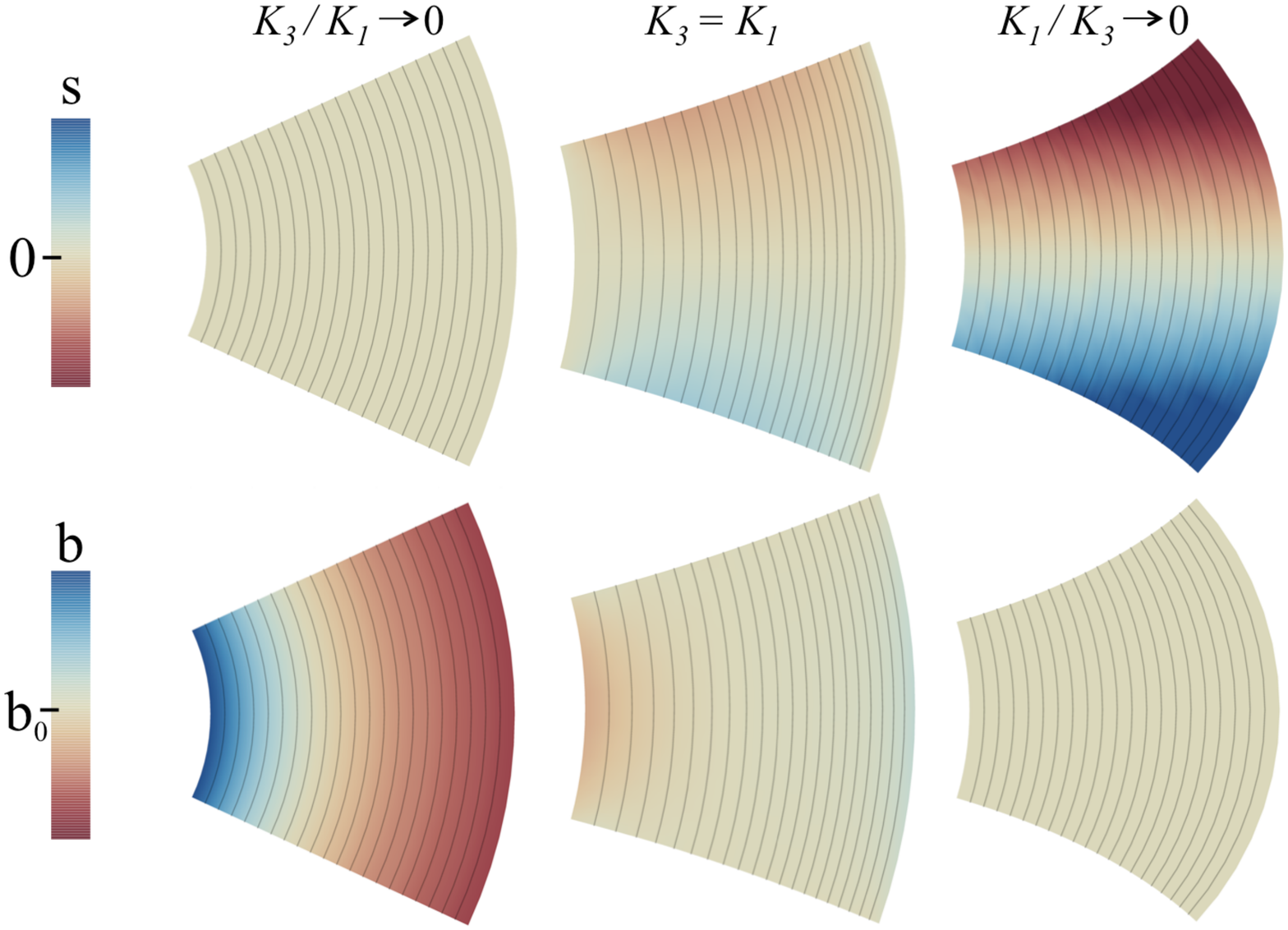}}
\caption{
Three limiting solutions for an attempted constant bend vanishing splay director in the plane. Zero splay (left panel), constant bend (right panel), and equal coupling constants (central panel). Top panels show the splay fields, and bottom panels show the bend field. Thin (black) lines denote the director integral curves.}\label{fig:splaybend}
\end{figure}

Due to its symmetry, the particular case of $K_{1}=K_{3}$, often termed the isotropic case, allows a particularly elegant solution. In this case the Euler Lagrange equation takes the form 
\[
\a s_{v}+\b b_{u}=0.
\]
Differentiating \eqref{eq:s&b2} yields
\[
\begin{aligned}
\a s_{v}=&\partial _{v}(\a s)- s \a_{v}=\partial_{u}\partial_{v} \log(\b)+s\, b\, \a\,\b ,\\
\b b_{u}=&\partial _{u}(\b b)- b \b_{u}=-\partial_{v}\partial_{u} \log(\a)-s\, b\, \a\,\b,
\end{aligned}
\]
thus producing for  the equilibrium condition $\partial_{u}\partial_{v} \log(\b/\a)=0$, for which the most general solution reads
\[
\a = A(u) \eta(u,v),\quad \b=B(v) \eta(u,v),
\]
for some three functions $A(u), \, B(v)$ and $\eta(u,v)$. We recall that any substitution $U(u)$ and $V(v)$ preserves the nature of the parametric curves as pointing along and perpendicularly to the director field, respectively. Through such a transformation we loose the gauge freedom we had in determining $\a$ and $\b$ along certain curves, but are allowed to set $A(u)=B(v)=1$. This implies that there exists a conformal map such that the director lines are the image of the cartesian  
x-parametric curves. The compatibility condition becomes 
\[
-\frac{\triangle \log(\eta)}{\eta^{2}}=K,
\]
while the splay and bend read
\[
s= \frac{\eta_u}{\eta^2}, \qquad
b= -\frac{\eta_v}{\eta^2}.
\]
Considering the Euclidean case in a rectangular domain and setting $\lambda=\log(\eta)$ we obtain that $\triangle \eta =0$ in the bulk, and use the boundary terms from the Euler Lagrange equations to close the boundary value problem. The central panel of Figure \ref{fig:splaybend} displays such a solution.

\section{Discussion}
In this work we identify the geometric charge that corresponds to the local intrinsic tendencies of the constituents of a frustrated two dimensional liquid crystal by providing an explicit formula for the Gaussian curvature associated with given splay and bend fields. Through this mapping we obtain access to many optimal embedding results that are formulated for different systems displaying two dimensional geometric frustration. 

\begin{figure}[h]
\centerline{\includegraphics[width=.35\textwidth]{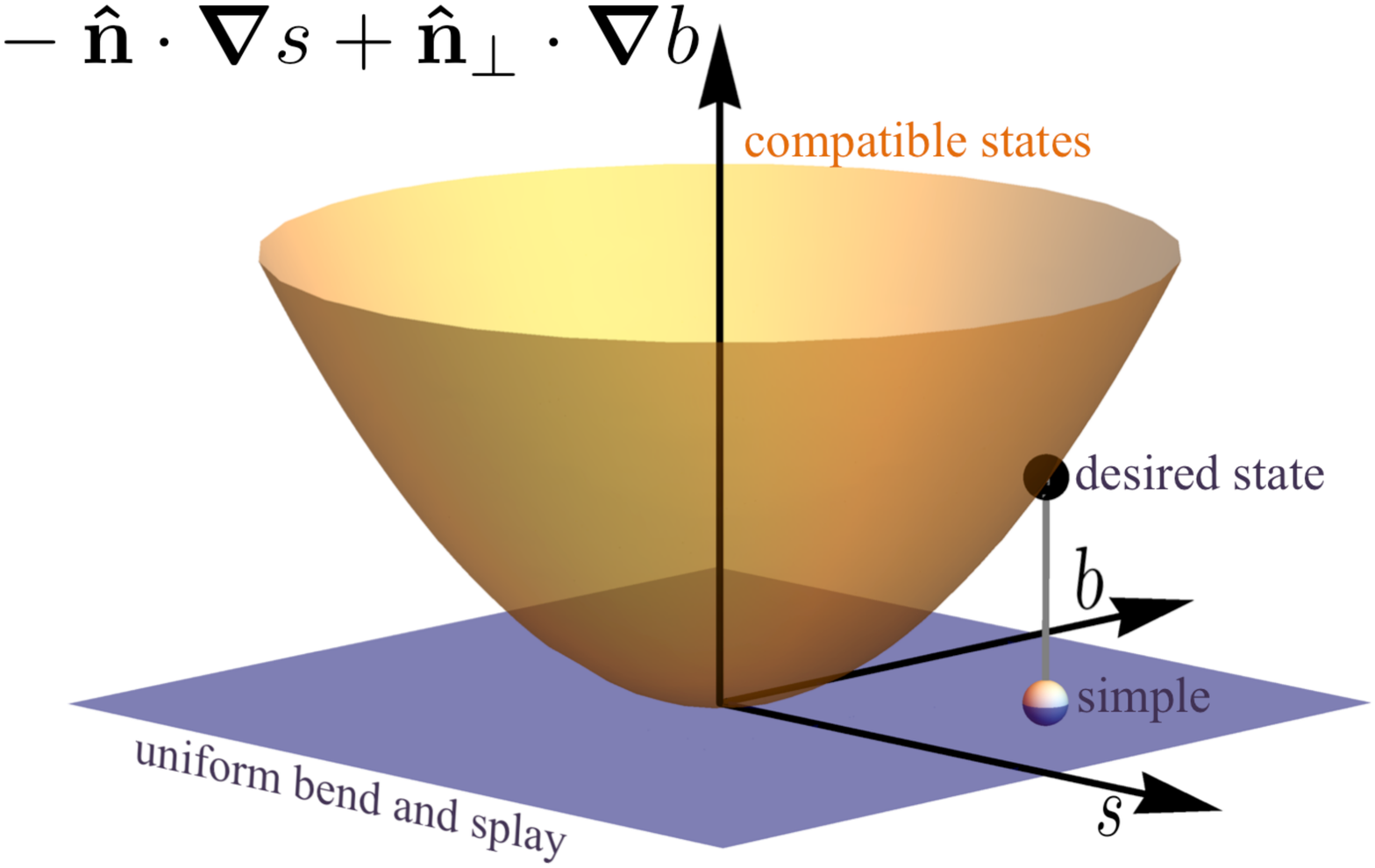}}
\caption{A three dimensional projection of the infinite dimensional space describing the local geometry of a 2D liquid crystalline phase. The splay, bend , and a combination of directed derivatives estimated at the center of a small square patch. The dimensions of the patch are smaller than the length scales associated with the bend and splay and their gradients, and the Frank energy coefficients are assumed equal. 
Uniform bend and splay fields are mapped to the horizontal (blue) plane, and the compatible phases are mapped to the (orange) paraboloid. A desired state (necessarily compatible) possessing gradients is marked by a black sphere. The constant bend and splay configuration marked by the light sphere is chosen such that the desired state will be the closest compatible state. 
}\label{fig:projection}
\end{figure}

Identifying the set of compatible intrinsic tendencies will also allow to harness the incompatibility in liquid crystals to produce simpler building blocks and more elaborate structures. In Figure \ref{fig:projection} a desired state characterized by spatial gradients of the splay and bend fields is achieved by uniform building blocks (that in particular favor no spatial gradients). The uniform phase, characterized by the non-zero constant bend and splay, favored by the building blocks is not compatible with the flat geometry, and the closest compatible state, the desired one, is realized instead. The Franck-free energy serves as a ``metric" for the configuration space determining which of the compatible states is closest to the attempted incompatible one.

We presented the compatibility conditions through both an Eulerian approach and a Lagrangean (intrinsic) approach. In the former the splay and bend fields are given in terms of the lab frame cartesian coordinates, i.e. $s(x,y)$ and $b(x,y)$, and the derivation of the compatibility conditions is more transparent. However, this type of question is artificial as the splay and bend fields are provided as functions of the embedding coordinates of the liquid crystalline phase, whose existence is sought through the compatibility conditions. It is thus more natural to consider the problem formulated in Lagrangean form where the attempted splay and bend fields are are provided as functions of a set of curvilinear coordinates that are oriented along and perpendicularly to the director. While such a description could be more directly related to the properties of the constituents of the liquid crystal, it renders the calculation itself more difficult. 
The intrinsic formulation may prove useful also to compatible systems as, for example, the Euclidean precompatibilty equation \ref{eq:flatprecompatibility} must hold for all director fields in the plane. Recent examples where the present formulation provides further insight into the geometry of the director include the theory of thin nematic elastomers \cite{ASK14,NS16}, and in the implementation of the tensorial conservation laws derived to describe nematic polymers \cite{SGP13}.

In many liquid crystalline systems, either boundary conditions or the topology of the system lead to non-trivial order. In the frustrated case, even free boundary problems result in non-trivial solutions that in particular depend on the relative value of the Frank coupling constants. The limit $K_{3}/K_{1}\to 0$ corresponds to the case where splay deviations are infinitely more energetically expensive compared to bending variations. We thus expect that in this limit, whenever possible, the system will completely comply with the local splay tendency and minimize the remaining bending term with respect to all such splay-free configurations. For bent core liquid crystals this implies that we seek to minimize $(b-\bar{b})^{2}$ with respect to all splay free solutions. A similar scenario occurs in the opposite limit where  $K_{1}/K_{3}\to 0$, and the bending tendency is obeyed. The numerical results for bent core liquid crystals support the results in these limits.
It is, however, a-priori unclear when could one assume the existence of such limits. Establishing the existence of these limits requires and a more detailed and formal mathematical treatment, and it outside the scope of this work.

Adequately describing the frustration of bent core liquid crystals in three dimensions and explaining the rich structures and strong optical activity such system display still evades us. Na\"ively one could argue that the results presented here support a scenario in which in order to simultaneously obey a zero splay and constant bend the nematogens organize into curved surfaces of negative Gaussian curvature $K=-\bar{b}^{2}$, that are then optimally packed to fill a volume. The optical activity in this case could be argued to be the outcome of the strong manifestation of handedness observed along asymptotic directions of hyperbolic surfaces, and predicted to locally average to zero if all directions are considered  \cite{EI14}.
However, this na\"ive approach not only ignores the non-trivial packing of surfaces of constant Gaussian curvature, but also underestimates the actual bend of the director in three dimensions. As discussed in section \ref{sec:Pre-comp-NE} the true bend in three dimensions $b^{3D}$ relates to the bend function used here, $b$, through $b^{3D}=\sqrt{b^{2}+k_{n}^{2}}$, where $k_{n}$ is the normal curvature of the surface along the direction of the director. Thus, mapping the three dimensional problem to a collection of two dimensional problems on curved surfaces requires some adaptations. However, we hope that the present work will lay the foundations for the geometric formulation of frustrated liquid crystals in three dimensions. 


\bibliography{bibfile}{}

\begin{thebibliography}{29}%
\makeatletter
\providecommand \@ifxundefined [1]{%
 \@ifx{#1\undefined}
}%
\providecommand \@ifnum [1]{%
 \ifnum #1\expandafter \@firstoftwo
 \else \expandafter \@secondoftwo
 \fi
}%
\providecommand \@ifx [1]{%
 \ifx #1\expandafter \@firstoftwo
 \else \expandafter \@secondoftwo
 \fi
}%
\providecommand \natexlab [1]{#1}%
\providecommand \enquote  [1]{``#1''}%
\providecommand \bibnamefont  [1]{#1}%
\providecommand \bibfnamefont [1]{#1}%
\providecommand \citenamefont [1]{#1}%
\providecommand \href@noop [0]{\@secondoftwo}%
\providecommand \href [0]{\begingroup \@sanitize@url \@href}%
\providecommand \@href[1]{\@@startlink{#1}\@@href}%
\providecommand \@@href[1]{\endgroup#1\@@endlink}%
\providecommand \@sanitize@url [0]{\catcode `\\12\catcode `\$12\catcode
  `\&12\catcode `\#12\catcode `\^12\catcode `\_12\catcode `\%12\relax}%
\providecommand \@@startlink[1]{}%
\providecommand \@@endlink[0]{}%
\providecommand \url  [0]{\begingroup\@sanitize@url \@url }%
\providecommand \@url [1]{\endgroup\@href {#1}{\urlprefix }}%
\providecommand \urlprefix  [0]{URL }%
\providecommand \Eprint [0]{\href }%
\providecommand \doibase [0]{http://dx.doi.org/}%
\providecommand \selectlanguage [0]{\@gobble}%
\providecommand \bibinfo  [0]{\@secondoftwo}%
\providecommand \bibfield  [0]{\@secondoftwo}%
\providecommand \translation [1]{[#1]}%
\providecommand \BibitemOpen [0]{}%
\providecommand \bibitemStop [0]{}%
\providecommand \bibitemNoStop [0]{.\EOS\space}%
\providecommand \EOS [0]{\spacefactor3000\relax}%
\providecommand \BibitemShut  [1]{\csname bibitem#1\endcsname}%
\let\auto@bib@innerbib\@empty
\bibitem [{\citenamefont {Onsager}(1949)}]{Ons49}%
  \BibitemOpen
  \bibfield  {author} {\bibinfo {author} {\bibfnamefont {L.}~\bibnamefont
  {Onsager}},\ }\href {\doibase 10.1111/j.1749-6632.1949.tb27296.x} {\bibfield
  {journal} {\bibinfo  {journal} {Annals of the New York Academy of Sciences}\
  }\textbf {\bibinfo {volume} {51}},\ \bibinfo {pages} {627} (\bibinfo {year}
  {1949})}\BibitemShut {NoStop}%
\bibitem [{\citenamefont {Meyer}(1976)}]{Mey76}%
  \BibitemOpen
  \bibfield  {author} {\bibinfo {author} {\bibfnamefont {R.~B.}\ \bibnamefont
  {Meyer}},\ }in\ \href@noop {} {\emph {\bibinfo {booktitle} {Molecular
  {{Fluids Les Houches Lectures}} 1973}}},\ \bibinfo {editor} {edited by\
  \bibinfo {editor} {\bibfnamefont {R.}~\bibnamefont {Balian}}\ and\ \bibinfo
  {editor} {\bibfnamefont {G.}~\bibnamefont {Weill}}}\ (\bibinfo  {publisher}
  {{Routledge}},\ \bibinfo {address} {London; New York; Paris},\ \bibinfo
  {year} {1976})\BibitemShut {NoStop}%
\bibitem [{\citenamefont {Grason}(2016)}]{Gra16}%
  \BibitemOpen
  \bibfield  {author} {\bibinfo {author} {\bibfnamefont {G.~M.}\ \bibnamefont
  {Grason}},\ }\href {\doibase 10.1063/1.4962629} {\bibfield  {journal}
  {\bibinfo  {journal} {The Journal of Chemical Physics}\ }\textbf {\bibinfo
  {volume} {145}},\ \bibinfo {pages} {110901} (\bibinfo {year}
  {2016})}\BibitemShut {NoStop}%
\bibitem [{\citenamefont {Kamien}\ and\ \citenamefont {Selinger}(2001)}]{KS01}%
  \BibitemOpen
  \bibfield  {author} {\bibinfo {author} {\bibfnamefont {R.~D.}\ \bibnamefont
  {Kamien}}\ and\ \bibinfo {author} {\bibfnamefont {J.~V.}\ \bibnamefont
  {Selinger}},\ }\href {\doibase 10.1088/0953-8984/13/3/201} {\bibfield
  {journal} {\bibinfo  {journal} {Journal of Physics: Condensed Matter}\
  }\textbf {\bibinfo {volume} {13}},\ \bibinfo {pages} {R1} (\bibinfo {year}
  {2001})}\BibitemShut {NoStop}%
\bibitem [{\citenamefont {Levin}\ and\ \citenamefont {Sharon}(2016)}]{LS16}%
  \BibitemOpen
  \bibfield  {author} {\bibinfo {author} {\bibfnamefont {I.}~\bibnamefont
  {Levin}}\ and\ \bibinfo {author} {\bibfnamefont {E.}~\bibnamefont {Sharon}},\
  }\href {\doibase 10.1103/PhysRevLett.116.035502} {\bibfield  {journal}
  {\bibinfo  {journal} {Physical Review Letters}\ }\textbf {\bibinfo {volume}
  {116}},\ \bibinfo {pages} {035502} (\bibinfo {year} {2016})}\BibitemShut
  {NoStop}%
\bibitem [{Note1()}]{Note1}%
  \BibitemOpen
  \bibinfo {note} {See equation \protect \textup {\hbox {\mathsurround \z@
  \protect \normalfont (\ignorespaces \ref {eq:FrankConstBend}\unskip
  \@@italiccorr )}} and the 2D polar limit of \cite {PSB+16}.}\BibitemShut
  {Stop}%
\bibitem [{\citenamefont {Lubensky}\ and\ \citenamefont
  {Radzihovsky}(2002)}]{LR02}%
  \BibitemOpen
  \bibfield  {author} {\bibinfo {author} {\bibfnamefont {T.~C.}\ \bibnamefont
  {Lubensky}}\ and\ \bibinfo {author} {\bibfnamefont {L.}~\bibnamefont
  {Radzihovsky}},\ }\href {\doibase 10.1103/PhysRevE.66.031704} {\bibfield
  {journal} {\bibinfo  {journal} {Physical Review E}\ }\textbf {\bibinfo
  {volume} {66}},\ \bibinfo {pages} {031704} (\bibinfo {year}
  {2002})}\BibitemShut {NoStop}%
\bibitem [{\citenamefont {Takezoe}\ and\ \citenamefont
  {Takanishi}(2006)}]{TT06}%
  \BibitemOpen
  \bibfield  {author} {\bibinfo {author} {\bibfnamefont {H.}~\bibnamefont
  {Takezoe}}\ and\ \bibinfo {author} {\bibfnamefont {Y.}~\bibnamefont
  {Takanishi}},\ }\href {\doibase 10.1143/JJAP.45.597} {\bibfield  {journal}
  {\bibinfo  {journal} {Japanese Journal of Applied Physics}\ }\textbf
  {\bibinfo {volume} {45}},\ \bibinfo {pages} {597} (\bibinfo {year}
  {2006})}\BibitemShut {NoStop}%
\bibitem [{\citenamefont {Shamid}\ \emph {et~al.}(2013)\citenamefont {Shamid},
  \citenamefont {Dhakal},\ and\ \citenamefont {Selinger}}]{SDS13}%
  \BibitemOpen
  \bibfield  {author} {\bibinfo {author} {\bibfnamefont {S.~M.}\ \bibnamefont
  {Shamid}}, \bibinfo {author} {\bibfnamefont {S.}~\bibnamefont {Dhakal}}, \
  and\ \bibinfo {author} {\bibfnamefont {J.~V.}\ \bibnamefont {Selinger}},\
  }\href {\doibase 10.1103/PhysRevE.87.052503} {\bibfield  {journal} {\bibinfo
  {journal} {Physical Review E}\ }\textbf {\bibinfo {volume} {87}},\ \bibinfo
  {pages} {052503} (\bibinfo {year} {2013})}\BibitemShut {NoStop}%
\bibitem [{\citenamefont {Gibaud}\ \emph {et~al.}(2012)\citenamefont {Gibaud},
  \citenamefont {Barry}, \citenamefont {Zakhary}, \citenamefont {Henglin},
  \citenamefont {Ward}, \citenamefont {Yang}, \citenamefont {Berciu},
  \citenamefont {Oldenbourg}, \citenamefont {Hagan}, \citenamefont {Nicastro},
  \citenamefont {Meyer},\ and\ \citenamefont {Dogic}}]{GBZ+12}%
  \BibitemOpen
  \bibfield  {author} {\bibinfo {author} {\bibfnamefont {T.}~\bibnamefont
  {Gibaud}}, \bibinfo {author} {\bibfnamefont {E.}~\bibnamefont {Barry}},
  \bibinfo {author} {\bibfnamefont {M.~J.}\ \bibnamefont {Zakhary}}, \bibinfo
  {author} {\bibfnamefont {M.}~\bibnamefont {Henglin}}, \bibinfo {author}
  {\bibfnamefont {A.}~\bibnamefont {Ward}}, \bibinfo {author} {\bibfnamefont
  {Y.}~\bibnamefont {Yang}}, \bibinfo {author} {\bibfnamefont {C.}~\bibnamefont
  {Berciu}}, \bibinfo {author} {\bibfnamefont {R.}~\bibnamefont {Oldenbourg}},
  \bibinfo {author} {\bibfnamefont {M.~F.}\ \bibnamefont {Hagan}}, \bibinfo
  {author} {\bibfnamefont {D.}~\bibnamefont {Nicastro}}, \bibinfo {author}
  {\bibfnamefont {R.~B.}\ \bibnamefont {Meyer}}, \ and\ \bibinfo {author}
  {\bibfnamefont {Z.}~\bibnamefont {Dogic}},\ }\href {\doibase
  10.1038/nature10769} {\bibfield  {journal} {\bibinfo  {journal} {Nature}\
  }\textbf {\bibinfo {volume} {481}},\ \bibinfo {pages} {348} (\bibinfo {year}
  {2012})}\BibitemShut {NoStop}%
\bibitem [{\citenamefont {Meng}\ \emph {et~al.}(2014)\citenamefont {Meng},
  \citenamefont {Paulose}, \citenamefont {Nelson},\ and\ \citenamefont
  {Manoharan}}]{MPNM14}%
  \BibitemOpen
  \bibfield  {author} {\bibinfo {author} {\bibfnamefont {G.}~\bibnamefont
  {Meng}}, \bibinfo {author} {\bibfnamefont {J.}~\bibnamefont {Paulose}},
  \bibinfo {author} {\bibfnamefont {D.~R.}\ \bibnamefont {Nelson}}, \ and\
  \bibinfo {author} {\bibfnamefont {V.~N.}\ \bibnamefont {Manoharan}},\ }\href
  {\doibase 10.1126/science.1244827} {\bibfield  {journal} {\bibinfo  {journal}
  {Science}\ }\textbf {\bibinfo {volume} {343}},\ \bibinfo {pages} {634}
  (\bibinfo {year} {2014})}\BibitemShut {NoStop}%
\bibitem [{\citenamefont {Schneider}\ and\ \citenamefont
  {Gompper}(2005)}]{SG05}%
  \BibitemOpen
  \bibfield  {author} {\bibinfo {author} {\bibfnamefont {S.}~\bibnamefont
  {Schneider}}\ and\ \bibinfo {author} {\bibfnamefont {G.}~\bibnamefont
  {Gompper}},\ }\href {\doibase 10.1209/epl/i2004-10464-2} {\bibfield
  {journal} {\bibinfo  {journal} {EPL (Europhysics Letters)}\ }\textbf
  {\bibinfo {volume} {70}},\ \bibinfo {pages} {136} (\bibinfo {year}
  {2005})}\BibitemShut {NoStop}%
\bibitem [{\citenamefont {Efrati}\ \emph {et~al.}(2013)\citenamefont {Efrati},
  \citenamefont {Sharon},\ and\ \citenamefont {Kupferman}}]{ESK13}%
  \BibitemOpen
  \bibfield  {author} {\bibinfo {author} {\bibfnamefont {E.}~\bibnamefont
  {Efrati}}, \bibinfo {author} {\bibfnamefont {E.}~\bibnamefont {Sharon}}, \
  and\ \bibinfo {author} {\bibfnamefont {R.}~\bibnamefont {Kupferman}},\ }\href
  {\doibase 10.1039/c3sm50660f} {\bibfield  {journal} {\bibinfo  {journal}
  {Soft Matter}\ }\textbf {\bibinfo {volume} {9}},\ \bibinfo {pages} {8187}
  (\bibinfo {year} {2013})}\BibitemShut {NoStop}%
\bibitem [{\citenamefont {Efrati}\ \emph {et~al.}(2009)\citenamefont {Efrati},
  \citenamefont {Sharon},\ and\ \citenamefont {Kupferman}}]{ESK09}%
  \BibitemOpen
  \bibfield  {author} {\bibinfo {author} {\bibfnamefont {E.}~\bibnamefont
  {Efrati}}, \bibinfo {author} {\bibfnamefont {E.}~\bibnamefont {Sharon}}, \
  and\ \bibinfo {author} {\bibfnamefont {R.}~\bibnamefont {Kupferman}},\ }\href
  {\doibase 10.1016/j.jmps.2008.12.004} {\bibfield  {journal} {\bibinfo
  {journal} {Journal of the Mechanics and Physics of Solids}\ }\textbf
  {\bibinfo {volume} {57}},\ \bibinfo {pages} {762} (\bibinfo {year}
  {2009})}\BibitemShut {NoStop}%
\bibitem [{\citenamefont {Armon}\ \emph {et~al.}(2011)\citenamefont {Armon},
  \citenamefont {Efrati}, \citenamefont {Kupferman},\ and\ \citenamefont
  {Sharon}}]{AEKS11}%
  \BibitemOpen
  \bibfield  {author} {\bibinfo {author} {\bibfnamefont {S.}~\bibnamefont
  {Armon}}, \bibinfo {author} {\bibfnamefont {E.}~\bibnamefont {Efrati}},
  \bibinfo {author} {\bibfnamefont {R.}~\bibnamefont {Kupferman}}, \ and\
  \bibinfo {author} {\bibfnamefont {E.}~\bibnamefont {Sharon}},\ }\href@noop {}
  {\bibfield  {journal} {\bibinfo  {journal} {Science}\ }\textbf {\bibinfo
  {volume} {333}},\ \bibinfo {pages} {1726} (\bibinfo {year}
  {2011})}\BibitemShut {NoStop}%
\bibitem [{\citenamefont {Hall}\ \emph {et~al.}(2016)\citenamefont {Hall},
  \citenamefont {Bruss}, \citenamefont {Barone},\ and\ \citenamefont
  {Grason}}]{HBBG16}%
  \BibitemOpen
  \bibfield  {author} {\bibinfo {author} {\bibfnamefont {D.~M.}\ \bibnamefont
  {Hall}}, \bibinfo {author} {\bibfnamefont {I.~R.}\ \bibnamefont {Bruss}},
  \bibinfo {author} {\bibfnamefont {J.~R.}\ \bibnamefont {Barone}}, \ and\
  \bibinfo {author} {\bibfnamefont {G.~M.}\ \bibnamefont {Grason}},\ }\href
  {\doibase 10.1038/nmat4598} {\bibfield  {journal} {\bibinfo  {journal}
  {Nature Materials}\ }\textbf {\bibinfo {volume} {15}},\ \bibinfo {pages}
  {727} (\bibinfo {year} {2016})}\BibitemShut {NoStop}%
\bibitem [{\citenamefont {Bruss}\ and\ \citenamefont {Grason}(2012)}]{BG12}%
  \BibitemOpen
  \bibfield  {author} {\bibinfo {author} {\bibfnamefont {I.~R.}\ \bibnamefont
  {Bruss}}\ and\ \bibinfo {author} {\bibfnamefont {G.~M.}\ \bibnamefont
  {Grason}},\ }\href {\doibase 10.1073/pnas.1205606109} {\bibfield  {journal}
  {\bibinfo  {journal} {Proceedings of the National Academy of Sciences}\
  }\textbf {\bibinfo {volume} {109}},\ \bibinfo {pages} {10781} (\bibinfo
  {year} {2012})}\BibitemShut {NoStop}%
\bibitem [{\citenamefont {Struik}(1988)}]{Str88}%
  \BibitemOpen
  \bibfield  {author} {\bibinfo {author} {\bibfnamefont {D.~J.}\ \bibnamefont
  {Struik}},\ }\href@noop {} {\emph {\bibinfo {title} {Lectures on classical
  differential geometry}}},\ \bibinfo {edition} {2nd}\ ed.,\ Dover books on
  advanced mathematics\ (\bibinfo  {publisher} {{Dover Publ}},\ \bibinfo
  {address} {New York},\ \bibinfo {year} {1988})\BibitemShut {NoStop}%
\bibitem [{Note2()}]{Note2}%
  \BibitemOpen
  \bibinfo {note} {If the splay and bend fields satisfy $\protect \boldsymbol
  {\nabla }s+\protect \mathrm {J}\protect \boldsymbol {\nabla }b=0$, the vector
  $\protect \mathbf {V}$ cannot be defined. However, for this case we obtain
  that $K=-s^2-b^2$ identically, which for constant Gaussian curvature can be
  easily shown to lead only to the uniform solution case. See appendix \ref
  {Appendix:Vis0}.}\BibitemShut {Stop}%
\bibitem [{\citenamefont {Bhatia}\ \emph {et~al.}(2013)\citenamefont {Bhatia},
  \citenamefont {Norgard}, \citenamefont {Pascucci},\ and\ \citenamefont
  {Bremer}}]{Bha13}%
  \BibitemOpen
  \bibfield  {author} {\bibinfo {author} {\bibfnamefont {H.}~\bibnamefont
  {Bhatia}}, \bibinfo {author} {\bibfnamefont {G.}~\bibnamefont {Norgard}},
  \bibinfo {author} {\bibfnamefont {V.}~\bibnamefont {Pascucci}}, \ and\
  \bibinfo {author} {\bibfnamefont {P.-T.}\ \bibnamefont {Bremer}},\ }\href
  {\doibase 10.1109/TVCG.2012.316} {\bibfield  {journal} {\bibinfo  {journal}
  {IEEE Transactions on Visualization and Computer Graphics}\ }\textbf
  {\bibinfo {volume} {19}},\ \bibinfo {pages} {1386} (\bibinfo {year}
  {2013})}\BibitemShut {NoStop}%
\bibitem [{Note3()}]{Note3}%
  \BibitemOpen
  \bibinfo {note} {The above elimination assumes that either $s_{u}\not =0$ or
  $b_{v}\not =0$. The where case both vanish corresponds to the $\protect
  \mathbf {V}=0$ discussed earlier and in appendix \ref
  {Apendix:Vis0}}\BibitemShut {NoStop}%
\bibitem [{Note4()}]{Note4}%
  \BibitemOpen
  \bibinfo {note} {This expression for the energy can be deduced by taking the
  two dimensional and polarized limit of the polar nematic theory of bent core
  liquid crystals, as for example appears in \cite {PSB+16}, and identifying
  the polar vector with the perpendicular to the director, $\protect \mathbf
  {p}\parallel \protect \mathbf {\protect \mathaccentV {hat}05E{n}}_{\perp }$.
  For non-positively defined $b$, it is natural (albeit less convenient) to
  formulate the bend term using squares: $(b^{2}-\protect \mathaccentV
  {bar}016{b}^{2})^{2}$.\cite {leo}}\BibitemShut {NoStop}%
\bibitem [{\citenamefont {Struik}(2012)}]{Str12}%
  \BibitemOpen
  \bibfield  {author} {\bibinfo {author} {\bibfnamefont {D.}~\bibnamefont
  {Struik}},\ }\href@noop {} {\emph {\bibinfo {title} {Lectures on {{Classical
  Differential Geometry}}: {{Second Edition}}}}},\ Dover Books on Mathematics\
  (\bibinfo  {publisher} {{Dover Publications}},\ \bibinfo {year}
  {2012})\BibitemShut {NoStop}%
\bibitem [{bot()}]{bote2}%
  \BibitemOpen
  \href@noop {} {\ }\BibitemShut {NoStop}%
\bibitem [{\citenamefont {Aharoni}\ \emph {et~al.}(2014)\citenamefont
  {Aharoni}, \citenamefont {Sharon},\ and\ \citenamefont {Kupferman}}]{ASK14}%
  \BibitemOpen
  \bibfield  {author} {\bibinfo {author} {\bibfnamefont {H.}~\bibnamefont
  {Aharoni}}, \bibinfo {author} {\bibfnamefont {E.}~\bibnamefont {Sharon}}, \
  and\ \bibinfo {author} {\bibfnamefont {R.}~\bibnamefont {Kupferman}},\ }\href
  {\doibase 10.1103/PhysRevLett.113.257801} {\bibfield  {journal} {\bibinfo
  {journal} {Physical Review Letters}\ }\textbf {\bibinfo {volume} {113}},\
  \bibinfo {pages} {257801} (\bibinfo {year} {2014})}\BibitemShut {NoStop}%
\bibitem [{\citenamefont {Nguyen}\ and\ \citenamefont {Selinger}(2016)}]{NS16}%
  \BibitemOpen
  \bibfield  {author} {\bibinfo {author} {\bibfnamefont {T.-S.}\ \bibnamefont
  {Nguyen}}\ and\ \bibinfo {author} {\bibfnamefont {J.~V.}\ \bibnamefont
  {Selinger}},\ }\href@noop {} {\bibfield  {journal} {\bibinfo  {journal}
  {arXiv:1612.06486 [cond-mat]}\ } (\bibinfo {year} {2016})}\BibitemShut
  {NoStop}%
\bibitem [{\citenamefont {Sven¨ek}\ \emph {et~al.}(2013)\citenamefont
  {Sven¨ek}, \citenamefont {Grason},\ and\ \citenamefont {Podgornik}}]{SGP13}%
  \BibitemOpen
  \bibfield  {author} {\bibinfo {author} {\bibfnamefont {D.}~\bibnamefont
  {Sven¨ek}}, \bibinfo {author} {\bibfnamefont {G.~M.}\ \bibnamefont {Grason}},
  \ and\ \bibinfo {author} {\bibfnamefont {R.}~\bibnamefont {Podgornik}},\
  }\href {\doibase 10.1103/PhysRevE.88.052603} {\bibfield  {journal} {\bibinfo
  {journal} {Physical Review E}\ }\textbf {\bibinfo {volume} {88}},\ \bibinfo
  {pages} {052603} (\bibinfo {year} {2013})}\BibitemShut {NoStop}%
\bibitem [{\citenamefont {Efrati}\ and\ \citenamefont {Irvine}(2014)}]{EI14}%
  \BibitemOpen
  \bibfield  {author} {\bibinfo {author} {\bibfnamefont {E.}~\bibnamefont
  {Efrati}}\ and\ \bibinfo {author} {\bibfnamefont {W.~T.}\ \bibnamefont
  {Irvine}},\ }\href@noop {} {\bibfield  {journal} {\bibinfo  {journal}
  {Physical Review X}\ }\textbf {\bibinfo {volume} {4}},\ \bibinfo {pages}
  {011003} (\bibinfo {year} {2014})}\BibitemShut {NoStop}%
\bibitem [{\citenamefont {Parsouzi}\ \emph {et~al.}(2016)\citenamefont
  {Parsouzi}, \citenamefont {Shamid}, \citenamefont {Borshch}, \citenamefont
  {Challa}, \citenamefont {Baldwin}, \citenamefont {Tamba}, \citenamefont
  {Welch}, \citenamefont {Mehl}, \citenamefont {Gleeson}, \citenamefont
  {Jakli}, \citenamefont {Lavrentovich}, \citenamefont {Allender},
  \citenamefont {Selinger},\ and\ \citenamefont {Sprunt}}]{PSB+16}%
  \BibitemOpen
  \bibfield  {author} {\bibinfo {author} {\bibfnamefont {Z.}~\bibnamefont
  {Parsouzi}}, \bibinfo {author} {\bibfnamefont {S.~M.}\ \bibnamefont
  {Shamid}}, \bibinfo {author} {\bibfnamefont {V.}~\bibnamefont {Borshch}},
  \bibinfo {author} {\bibfnamefont {P.~K.}\ \bibnamefont {Challa}}, \bibinfo
  {author} {\bibfnamefont {A.~R.}\ \bibnamefont {Baldwin}}, \bibinfo {author}
  {\bibfnamefont {M.~G.}\ \bibnamefont {Tamba}}, \bibinfo {author}
  {\bibfnamefont {C.}~\bibnamefont {Welch}}, \bibinfo {author} {\bibfnamefont
  {G.~H.}\ \bibnamefont {Mehl}}, \bibinfo {author} {\bibfnamefont {J.~T.}\
  \bibnamefont {Gleeson}}, \bibinfo {author} {\bibfnamefont {A.}~\bibnamefont
  {Jakli}}, \bibinfo {author} {\bibfnamefont {O.~D.}\ \bibnamefont
  {Lavrentovich}}, \bibinfo {author} {\bibfnamefont {D.~W.}\ \bibnamefont
  {Allender}}, \bibinfo {author} {\bibfnamefont {J.~V.}\ \bibnamefont
  {Selinger}}, \ and\ \bibinfo {author} {\bibfnamefont {S.}~\bibnamefont
  {Sprunt}},\ }\href {\doibase 10.1103/PhysRevX.6.021041} {\bibfield  {journal}
  {\bibinfo  {journal} {Physical Review X}\ }\textbf {\bibinfo {volume} {6}},\
  \bibinfo {pages} {021041} (\bibinfo {year} {2016})}\BibitemShut {NoStop}%
\end{thebibliography}%

\appendix
\section{Director aligned coordinate system and precompatibility through Gauss Bonnet}
\label{app:coordinates}
\begin{figure}[h]
\centerline{\includegraphics[width=.25\textwidth]{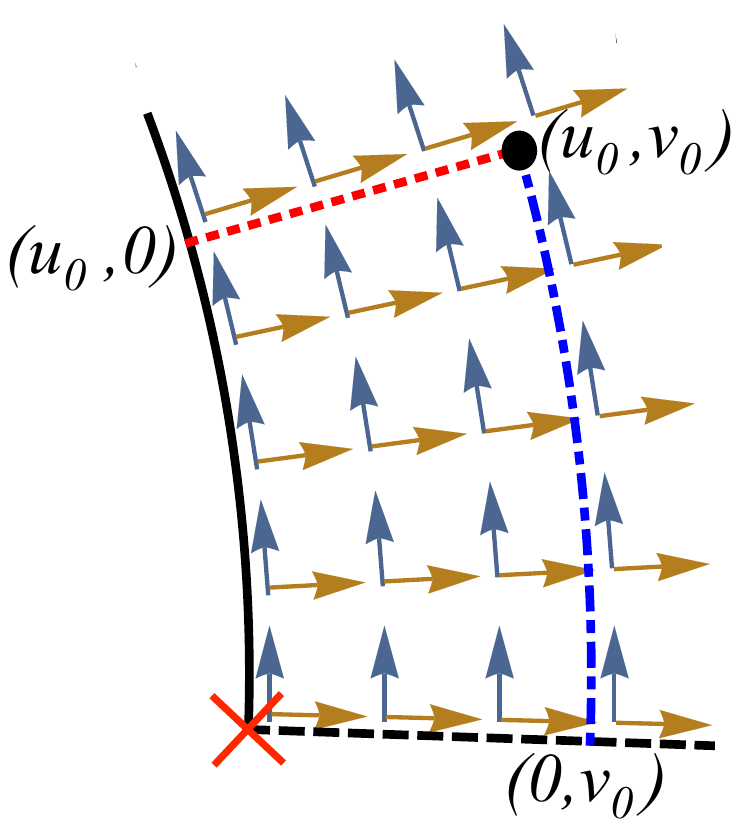}}
\caption{The origin, marked in (Red) cross, and base curves marked by (Black) dashed and solid curves. The (Blue) upright arrows denote the local orientation of the director field. 
}\label{fig:meshgrid}
\end{figure}

Give a surface $\mathcal{S}$ and a director field $\n$ on  $\mathcal{S}$ we come to construct a parametrization on  $\mathcal{S}$ such that the parametric curves are oriented parallel and perpendicular to the director field. We first choose an origin, a point on $\mathcal{S}$, that will be attributed the coordinates $u=0$ and $v=0$.
From the origin we construct the integral curve of the director filed and the integral curve of the normal to the director; these are termed the base curves and are marked by (black) solid and dashed curves in figure \ref{fig:meshgrid}. We parametrize the base curves by arclength; $u$ measures arclength along the director integral curve and $v$ measures arclength along the integral curve of the director's normal. Provided $s^{2}+b^{2}$ is bounded then we know that all ingetral curves in the domain do not curve too much, and we can find a domain on $\mathcal{S}$ such that for every point in the domain the director, and director's normal integral curves that start at the point, each cross one of the base curves. The point inherits its coordinates from the base curve's arclength coordinate at the point of crossing. For example, in Figure \ref{fig:meshgrid} we follow the director's integral curve (blue dash-dotted line) from the point until we cross the dashed base-line at the point $(0,v_{0})$. Following the director's normal integral curve (red dotted line) leads to the intersection with the second base curve at $(u_{0},0)$. The point is thus attributed the coordinates $(u_{0},v_{0})$. Note that keeping the $v$-coordinate constant and varying $u$ thus follows a director integral curve, and that correspondingly the $v$-coordinate parametric curve (along which $u$ is constant) are director's normal integral curves. Last we note that by construction here $\a(u,0)=1$ and $\b(0,v)=1$. However, One could reparametrize the base curve to other than arclength variables $V(v)$ and $U(u)$ while keeping the property that parametric curves are tangent and perpendicular to the local director. 

We can now formulate the precompatibility condition as an immediate corollary of the Gauss-Bonnet theorem:
We consider a closed curve composed of four parametric curves for example the trajectory 
\[
(0,0)\to(0,v_{0})\to(u_{0},v_{0})\to(u_{0},0)\to(0,0)
\]
 in Figure \ref{fig:meshgrid}. The Gauss Bonnet theorem reads:
\[
\iint K\,dA=-\oint k_g dl+\sum_i \theta_i-2\pi
\]
We note that the integral curves are everywhere perpendicular thus all angles are right angles, and that directly from the metric \ref{eq:metric} one can show the geodesic curvature of the director and director's normal integral curves read $k_{g}=-b$ and $k_{g}=-s$ respectively. Thus we obtain   
\[
\oint k_g dl=\oint\left(b\n-s\np\right)\cdot d\vec{l}.
\]
Substituting into the Gauss Bonnet theorem and using Stokes theorem we obtain
\[
\iint\left[ K+\bnabla\cdot\left(s\n-b\np\right)\right]dA=0.
\]
As this integral equation must hold for all (and in particular arbitrarily small) parametric quadrangles its integrand must vanish locally and we recover the precompatibility condition  \ref{eq:precompatibility}:
\begin{equation}
K=\bnabla\cdot\left(-s\n+b\np\right).
\end{equation}

\section{Sign ambiguity in the reconstruction formula}
We start by considering the compatible splay and bend fields $b=1$ and $s=-\frac{x}{\sqrt{1-x^2}}$. The splay and bend gradients produce constant unit vectors : $\hat{\mathbf{V}}=(-1,0)$ and $\hat{\mathbf{V}}_{\perp}=(0,-1)$, and the scalar 
\[
\frac{b^{2}+s^{2}+K}{\left|\bnabla s+\j\bnabla b\right|}=\sqrt{1-x^{2}}. 
\]
Substituting into \ref{eq:directorReconstruct} gives
\[
\n=(\sqrt{1-x^{2}},\mp |x|).
\]
We note that the correct values for the bend are obtained only if we use $\n_{y}=-|x|$ for $x<0$ and $\n_{y}=|x|$ for $x>0$. The resulting director 
 $\n=\left(\sqrt{1-x^2},x\right)$ is smooth, and recovers the splay and bend functions.

 \section{Euler-Lagrange and stability condition}
We assume that the given director at equilibrium corresponds to the splay and bend fields $s$ and $b$. These in turn minimize the Frank energy with respect to all possible director fields. Small variations of the director correspond to locally small rotations by a small angle $\theta$: $\n'=\cos(\theta)\n+\sin(\theta) \np$, where $\theta\ll 1$. Explicit substitution in equation \ref{eq:s&b} shows that such an infinitesimal rotation leads to variations in the splay and bend according to
\begin{align*}
s' &=s-b\,\theta+\frac{\theta_v}{\beta}  \\
b' &=b+s\,\theta+\frac{\theta_u}{\alpha}
\end{align*}
We now calculate the energy difference to first order in $\theta$ using the unperturbed coordinate system $u$, and $v$. Note, that in particular this implies that the $u$ direction will no longer point along the perturbed director, and that the area element remains unchanged. 
\begin{align}
\Delta E=2\!\! \int\!\! \Big[&K_1 s\left(\frac{\theta_v}{\beta}-b\theta\right)\nonumber\\
+&K_3\left(b-\bar{b}\right)\left(s\theta+\frac{\theta_u}{\alpha}\right)\Big]\alpha\beta\,du\,dv.
\end{align}
We first consider only differential rotations that vanish on the boundary of the region considered. Integrating by parts, and eliminating the boundary contribution leads to the following Euler-Lagrange equation:
\begin{equation}
\label{eq:minimumenergy}
K_1 \frac{s_v}{\beta}+K_3 \frac{b_u}{\alpha}=
K_1 \np\!\!\cdot\!\!\bnabla s+K_3 \n\!\cdot\!\!\bnabla b=0.
\end{equation}

\section{Application to compatible liquid crystal systems}
Recently, a geometric theory for thin nematic elastomer was presented \cite{ASK14}.  The theory studied a thin nematic elastomer that displays a length shrinkage by a factor $\lambda^{1/2}$ along the director and a factor $\lambda^{-\nu_{t}/2}$ in the perpendicular direction when activated. The work provided a direct formula relating the Gaussian curvature of the geometry induced by this differential shrinkage as a function of the directors angle and its first and second derivatives. As this quantity is clearly coordinate independent, it must allow an intrinsic formulation using the bend and splay of the unperturbed field. Exploiting equation \ref{eq:flatprecompatibility} we obtain the compact result
\[
\begin{aligned}
K_{shrinked}=&(\lambda-\lambda^{-\nu_{t}})\bigl( s^{2}+\n\cdot\bnabla s\bigr)\\
=&(\lambda-\lambda^{-\nu_{t}})\bigl(- b^{2}+\np\cdot\bnabla b\bigr),
\end{aligned}
\]
where the last equality holds because the initial director was cast in a flat geometry.

Another application can be found in the implementation of the tensorial conservation laws derived to describe nematic polymers \cite{SGP13}. In the limit of uniform density and nematic degree the authors obtain the equation
\[
\bnabla\cdot [\n (\bnabla\cdot \n)]+\bnabla\cdot [(\n \cdot\bnabla)\n)]=0.
\] 
Considering directors with no $z$ components we obtain an effectively planar two dimensional system, for which
the flat precompatibility equation assumes a similar form only with the opposite sign between the two terms. This implies that each of the terms above vanish independently; $s^{2}+\n\cdot \bnabla s=b^{2}-\np \cdot \bnabla b=0$.

\end{document}